\documentclass[sigplan,screen,authorversion,nonacm]{acmart} 
\AtBeginDocument{%
  }

\setcopyright{rightsretained}
\copyrightyear{2025}
\acmYear{2025}
\acmDOI{XXXXXXX.XXXXXXX}
\acmConference[ICOOOLPS '25]{ICOOOLPS '25}{June 30 -- July 4,
  2025}{Bergen, Norway}

\acmISBN{None}

\usepackage{listings}
\usepackage{xcolor}
\usepackage{tikz}
\DeclareRobustCommand*\circled[1]{\tikz[baseline=(char.base)]{
            \node[shape=circle,draw,inner sep=2pt] (char) {#1};}}

\usepackage{ifthen}

\newcommand{\anonurl}[1]{\anon[URL removed for anonymity reasons]{#1}}

\begin{document}

\title{Low Overhead Allocation Sampling in a Garbage Collected Virtual Machine}

\author{Christoph Jung}

\email{Christoph.Jung@hhu.de}
\orcid{0009-0004-9348-0468}
\affiliation{%
  \institution{Heinrich Heine University Düsseldorf}
  \city{Düsseldorf}
  \country{Germany}
}

\author{CF Bolz-Tereick}
\email{cfbolz@gmx.de}
\orcid{0000-0003-4562-1356}
\affiliation{%
  \institution{Heinrich Heine University Düsseldorf}
  \city{Düsseldorf}
  \country{Germany}
}


\begin{abstract}

Compared to the more commonly used time-based profiling, allocation profiling
provides an alternate view of the execution of allocation
heavy dynamically typed languages. However, profiling every single
allocation in a program is very inefficient. We present a sampling
allocation profiler that is deeply integrated into the garbage collector of
PyPy, a Python virtual machine. This integration ensures tunable low
overhead for the allocation profiler, which we measure and quantify. Enabling
allocation sampling profiling with a sampling period of 4 MB leads to a
maximum time overhead of 25\% in our benchmarks, over un-profiled regular
execution.
\end{abstract}


\maketitle

\section{Introduction}

There are many time-based statistical profilers for various programming
languages. They allow programmers to gain insights into where their software is
spending time, without causing too much overhead. Examples include perf (for
full-system and kernel profiling on Linux) and many language-specific tools
such as py-spy\footnote{\url{https://github.com/benfred/py-spy}}
and VMProf\footnote{\anonurl{\url{https://github.com/Cskorpion/vmprof-python/tree/pypy_gc_allocation_sampling_obj_info}}}
for Python.

On the other hand, there are memory profilers such as
Memray\footnote{\url{https://github.com/bloomberg/memray}} for Python
and VisualVM\footnote{\url{https://github.com/oracle/visualvm}} for Java. They can be handy for finding leaks or for
discovering functions that allocate a lot of memory. Memory profilers
typically profile every single allocation done. This results in precise
profiling but larger overhead.

In this paper we present our experimental approach to low-overhead statistical
memory profiling. Instead of instrumenting every allocation, we take a sample
every n-th allocated byte. This allows programmers and VM implementers to gain
information about where allocations happen. We have tightly integrated VMProf (PyPy's
time-based sampling profiler) and the 
PyPy\footnote{\anonurl{\url{https://github.com/Cskorpion/pypy/tree/gc_allocation_sampling_obj_info_u_2.7}}}
garbage collector to achieve this.

Our main technical insight is that the check whether an allocation should be
sampled can be made free.
This is done by folding it into the bump-pointer
allocator check that PyPy's GC uses to find out if it should start a minor
collection. In this way the fast path with and without memory sampling are
exactly the same.

Performing sampling on the level of the GC is also important because it gives
more accurate numbers compared to source- or bytecode-level instrumentation.
PyPy's JIT does escape analysis~\cite{BolzCF:allrembpeiatj} to remove a lot of
short-lived allocations. This means, that for example, instrumenting Python
bytecode to profile allocations would give the wrong picture.
On the other hand, the JIT also introduces a lot of new allocations for itself,
both during its optimization phase but also at run-time, because it uses
heap-allocated frames.

\begin{figure*}[]
  \centering
    \includegraphics[width=0.9\textwidth]{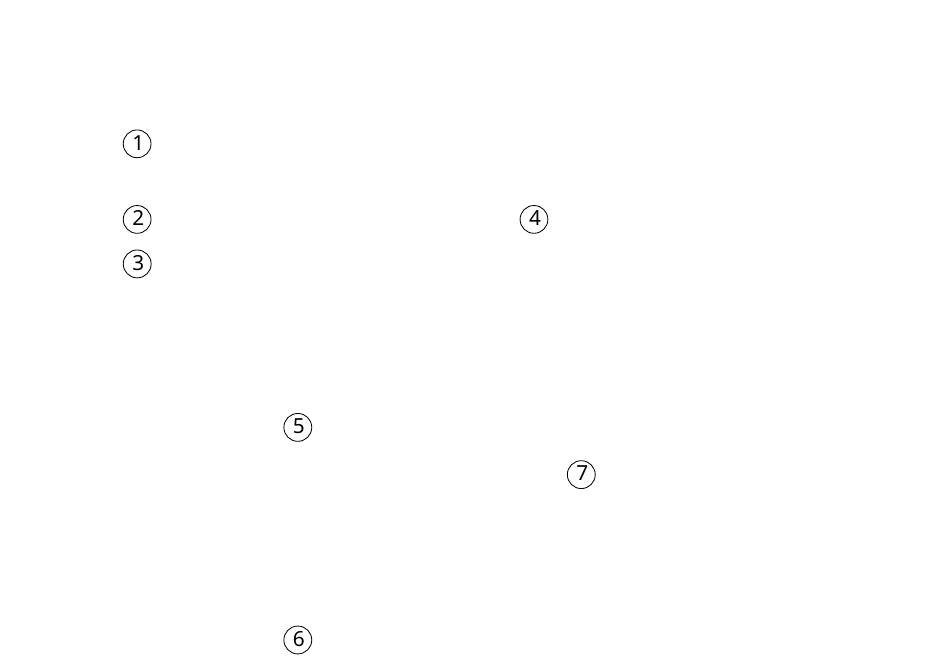}
    \caption{Firefox Profiler Call Tree View showing VMProf data. The three Memory tracks
    \circled{1} show different kinds of memory statistics. There are two tracks
    with stack samples, both allocation samples \circled{2} and time-based
    samples \circled{3}. The thin black vertical marks \circled{4} above the
    allocation sample track are the points where minor collections happen, the
    wider rectangles are (incremental) major collections.
    In the lower half, the call-tree pane is active,
    showing inverted call stacks. The top stack frames \circled{5} and
    \circled{6} show the type of object that triggered the corresponding
    allocation samples. The color of the type frames show how quickly the
    object died, if the type name has a green square \circled{5} the object died
    before it was tenured, if it's red \circled{6} it survived at least one
    minor collection. The frames below the top frame show \circled{7} show the
    stack of Python frames that caused the allocation.}
  \label{fig:firefox_view}
\end{figure*}

The profiler captures a call stack of the allocation site. Additionally, it
stores the type of the sampled objects and whether they got tenured to the old
generation after the next minor collection. We also
store general heap statistics like total live heap size of the GC and total RSS
of the process. The resulting profile can be visualized using the Firefox
Profiler user interface by converting our internal format into a JSON file that
the Firefox Profiler UI can present. This gives a convenient way to consume and
analyze the resulting data. An example profile can be seen in Figure~\ref{fig:firefox_view}.

The contributions of this paper are:

\begin{itemize}
    \item We integrate a sampling memory profiler tightly into PyPy's GC.
    \item We enrich the captured profiles by also storing the RPython-level
        types of the sampled objects, as well as their survival length.
    \item We convert\footnote{\anonurl{\url{https://github.com/Cskorpion/vmprof-firefox-converter/tree/allocation_sampling_obj_info}}}
          VMProf's internal format to that of the Firefox Profiler.
    \item We measure the overhead of sampling allocations.
\end{itemize}

All the work is open source, and we hope to contribute it to upstream PyPy and
VMProf in the future.

The paper is structured as follows: we give some background information about
the involved technologies in Section~\ref{sec:background}, then explain our
technical approach in Section~\ref{sec:approach}. In Section~\ref{sec:firefox}
we present our conversion tool that can turn the data our profiler produces into
a format that can be visualized with the Firefox Profiler UI. We evaluate the
overhead of sampling allocations in Section~\ref{sec:evaluation} and also
discuss a small example problem we identified with the memory profiler. Then we
discuss future ideas, related work, and finally conclude.

\section{Background}
\label{sec:background}

In this section we explain the details necessary to understand the technical
contributions of our paper, to make the writing self-contained. For more
general information about garbage collection we refer the reader to the relevant
literature~\cite{wilson_uniprocessor_1992, jones_garbage_2023}.

\subsection{PyPy and its GC}

PyPy\footnote{\url{https://pypy.org}} is an alternative Python implementation~\cite{rigo_pypys_2006}.
Written in RPython~\cite{AnconaD:rpystetrdastol}, it features a meta-tracing
just-in-time compiler~\cite{BolzCF:trametlptjc} that can give significant performance
gains for pure Python programs over the default CPython implementation.

Another major difference to CPython is that PyPy does not use reference
counting for memory management but a generational incremental collector.
That means, there are two spaces for allocated objects, the nursery and the
old-space. Freshly allocated objects will be bump-pointer allocated into the
nursery. When the nursery is full at some point, a minor collection is
performed. Then all
surviving objects will be tenured, i.e., moved into the old-space. The
old-space is much larger than the nursery and is collected less frequently and
incrementally, using a mark-and-sweep approach.

\subsection{Bump-Pointer Allocation in the Nursery}

The nursery is a small continuous memory area (typically a few megabytes in
size) that utilizes two pointers to
keep track of the start and end of the free space available for allocation in
it. They are called
\texttt{nursery\_free} and \texttt{nursery\_limit}. When memory is allocated, the
GC checks if there is enough space in the nursery left. If there is, the
\texttt{nursery\_free} pointer will be returned as the start address for the
newly allocated memory, and \texttt{nursery\_free} will be moved forward by the
amount of allocated memory. This is the fast path of allocation in a nursery, and
it is very efficient: if there is space in the nursery, allocation takes only a
couple of instructions. 
Note that there is only one nursery, into which every thread allocates.

\begin{figure}
  \centering
  \includegraphics[width=9cm]{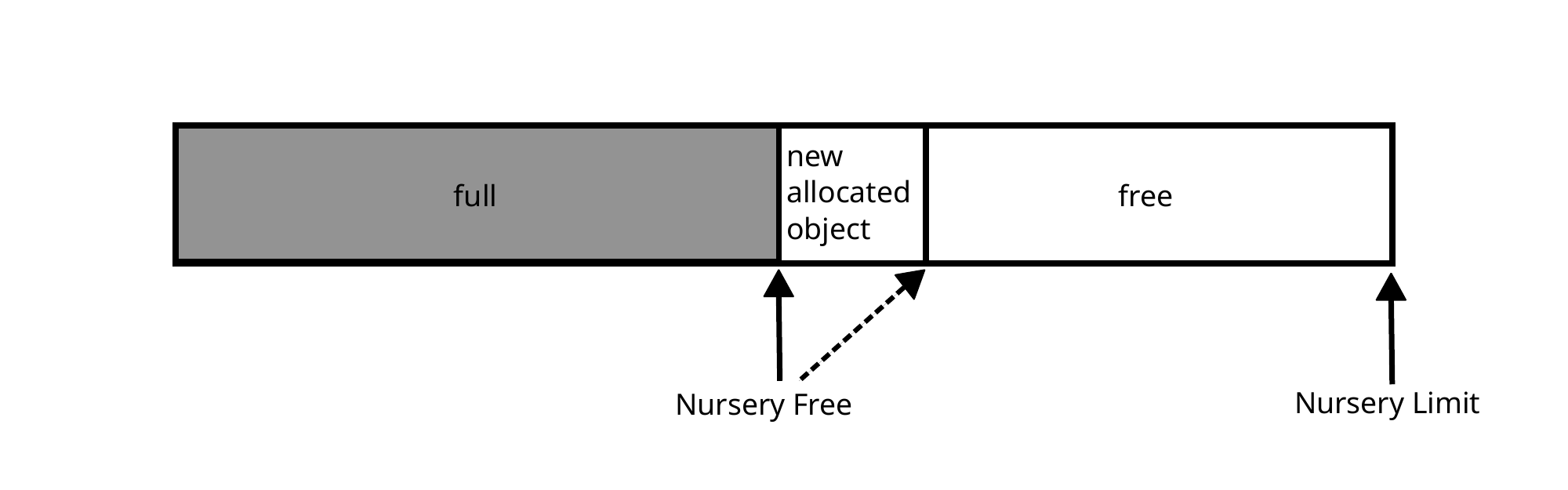}
  \caption{Nursery allocation, fast path taken}
  \label{fig:nursery_allocation}
\end{figure}

\begin{figure}
  \centering
  \includegraphics[width=9cm]{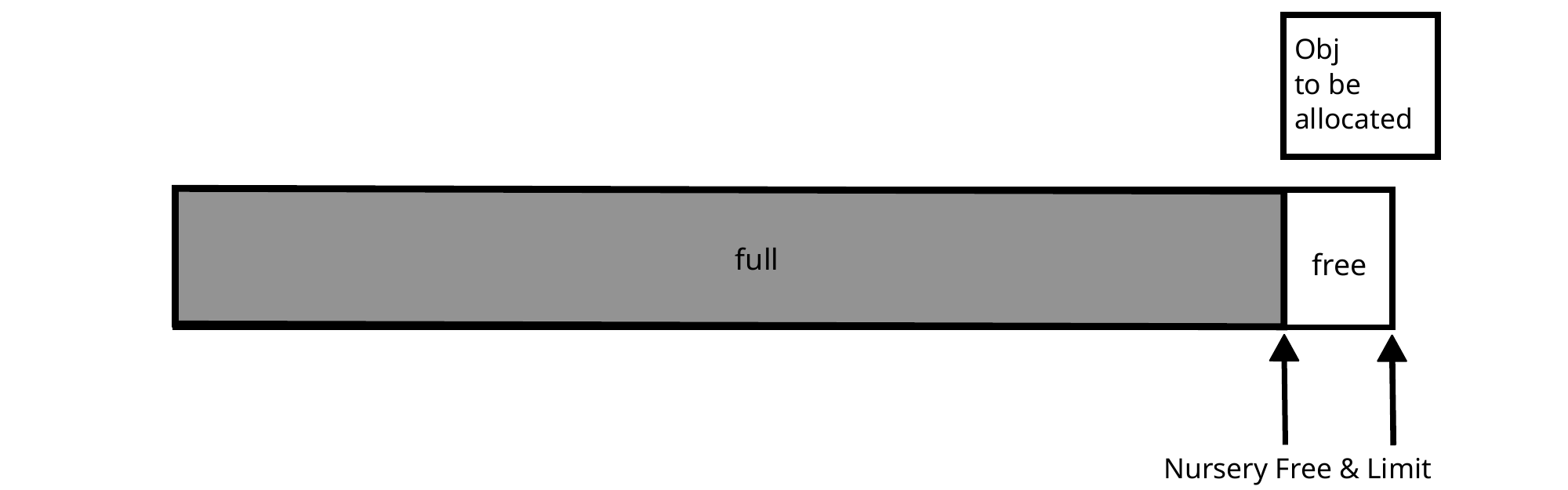}
  \includegraphics[width=9cm]{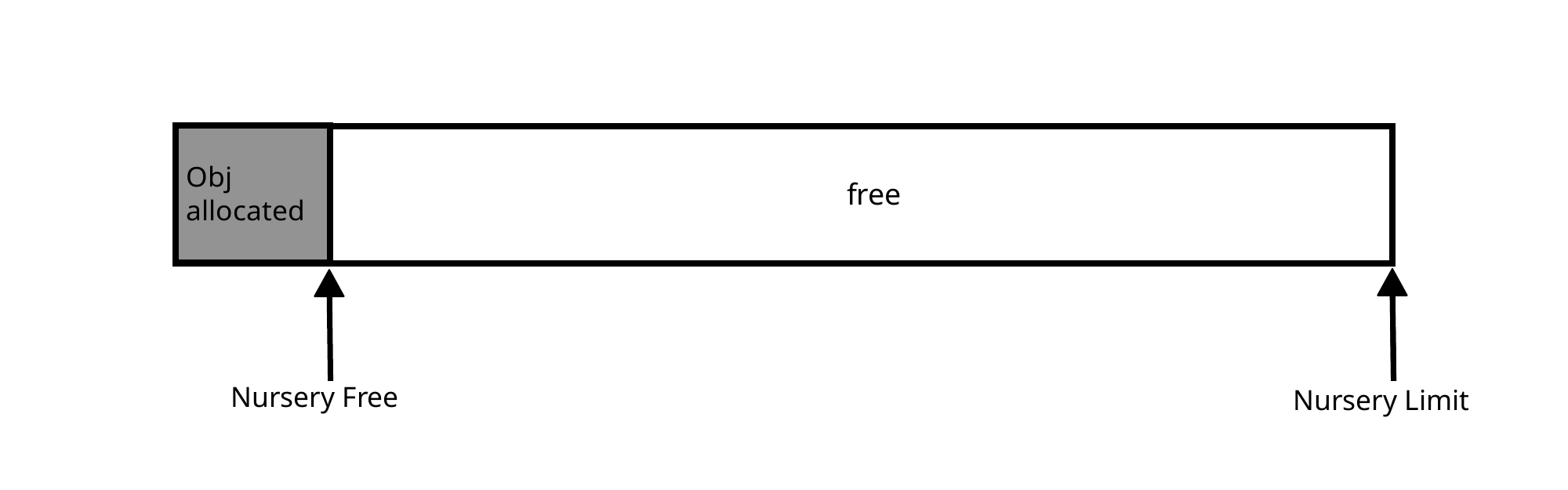}
    \caption{Nursery allocation, but the nursery is full (upper picture).
    Therefore, we perform a minor collection, which evacuates the nursery, and
    the newly allocated object can be placed at the beginning (lower picture).}
  \label{fig:nursery_full}
\end{figure}

At some point there won't be enough space in the nursery left to fulfill an
allocation request. Then \texttt{collect\_and\_re\-serve} is called to start a
minor collection and allocate afterwards. A nursery collection will move all
surviving objects into the old-space so that the nursery is empty, and the
requested allocation can be made. Figures~\ref{fig:nursery_allocation} (there
is enough space in the nursery) and~\ref{fig:nursery_full} (not enough space,
need to collect) illustrate this process. Nursery allocation is shown as
pseudocode in Figure~\ref{fig:allocate}.

Pseudo-code for \texttt{collect\_and\_reserve} is shown in
Figure~\ref{fig:collect_and_reserve_plain}. It triggers a minor collection and
sets the \texttt{nursery\_free} according to the allocation size and returns the
start of the nursery afterwards.

\begin{figure}
  \begin{lstlisting}
def allocate(size):
    # result is a pointer into the nursery, obj will be allocated there
    result = gc.nursery_free
    # Move nursery_free pointer forward by size
    gc.nursery_free = result + size
    # Check if this allocation exceeds the nursery
    if gc.nursery_free > gc.nursery_limit:
        # If it does => collect the nursery and allocate afterwards
        result = collect_and_reserve(size)
    return result
  \end{lstlisting}
  \caption{Pseudo-code for allocation function of PyPy's GC}
  \label{fig:allocate}
\end{figure}

\begin{figure}
  \begin{lstlisting}
def collect_and_reserve(size):
    gc.minor_collection()
    result = gc.nursery_free
    gc.nursery_free += size
    return result
  \end{lstlisting}
  \caption{Pseudo-code for collect\_and\_reserve}
  \label{fig:collect_and_reserve_plain}
\end{figure}

\subsection{Large-Object Allocations}
\label{section-exterenal-allocations}

The nursery is usually no more than a few megabytes in size. If an allocation
size exceeds a certain threshold, it will therefore be allocated in a separate
large-object space (which is collected with a mark-sweep approach).
Those objects are allocated into a separate memory area outside of the nursery and
old-space, thus don't use the nursery bump-pointer logic for allocation.

\subsection{VMProf}

VMProf\footnote{\url{https://vmprof.readthedocs.io/en/latest/}}
is a statistical time-based profiler for PyPy.
VMProf samples the call stack of the running Python code a user-configured number of times per second. 
By adjusting this number, the overhead of profiling can be modified to pick a
trade-off between overhead and precision of the profile. In the resulting
profile, functions that run for a long time stand out the most; functions with shorter
run time less so. VMProf was developed by PyPy developers for profiling PyPy, as
`normal' CPython profilers don't work well with PyPy, or have a lot of
overhead.

\subsection{Firefox Profiler UI}

The Firefox Profiler\footnote{\url{https://profiler.firefox.com/}}
is a tool for analyzing the performance of web-based applications.
Profiles from other profilers can be imported into the Firefox Profiler.
The UI features multiple ways of visualizing profiled information, like call
trees, flame graphs, stack charts, assembly- and source-code view, a zoomable
timeline, and more.
By using the vmprof-firefox-converter, it is possible to view VMProf profiles with the Firefox Profiler's user interface.

\section{Technical Details of our Approach}
\label{sec:approach}

\subsection{Sampling from a conceptual point of view}

We want to sample every n-th byte that is being allocated, which is configured by
the parameter \texttt{sample\_n\_bytes} by the user. This parameter is called
the \emph{sampling period}, the inverse of the sampling frequency. To do that, we want to keep
track of how many bytes have been allocated since the last sample in a new variable
\texttt{allocated} in the GC. Every time we
allocate, we add the number of bytes allocated to that number. If the number
\texttt{allocated} exceeds \texttt{sample\_n\_bytes},
we want to sample the allocation. To achieve this, we could change the
allocation logic to the one in Figure~\ref{fig:slow}.

\begin{figure}
  \begin{lstlisting}
def allocate_with_sample_slow(size):
    # result is a pointer into the nursery, obj will be allocated there
    result = gc.nursery_free
    # Move nursery_free pointer forward by size
    gc.nursery_free = result + size
    # Check if this allocation exceeds the nursery
    if gc.nursery_free > gc.nursery_limit:
        # If it does => collect the nursery and allocate afterwards
        result = collect_and_reserve(size)
    # new code:
    gc.allocated += size
    if gc.allocated >= gc.sample_n_bytes:
        vmprof.sample_now()
        gc.allocated = 0
    return result
  \end{lstlisting}
  \caption{Pseudo-code for an inefficient way to implement allocation sampling}
  \label{fig:slow}
\end{figure}

However, this approach has a big downside. Specifically, it will make the fast
path of allocation roughly twice as expensive, no matter how large
\texttt{sample\_n\_bytes} is. Given that allocation is an extremely common
operation in PyPy and given that it is currently very fast, we wanted to avoid
that. PyPy's GC manages to achieve a peak allocation rate of about 11 GB/s on 
the benchmark machine (see Section~\ref{sec:evaluation_overhead}).

To improve this, we can observe that there is already an addition of
\texttt{size} in the \texttt{allocate} function, and a
comparison with a limit, the \texttt{nursery\_limit}. We want to reuse these for
also keeping track of whether we need to sample or not.

\subsection{Sampling efficiently}
\label{sec:sampling_efficiently}

We want to reuse the check for whether the nursery is full in the allocation
fast path to perform a dual-function. The check is then used both to check
whether we should start a minor collection or whether we should take an
allocation sample.

Usually, when there is not enough space in the nursery left to fulfill an
allocation request, the nursery will be collected and the allocation will be
done afterwards. We reuse that mechanism for sampling by introducing a new
pointer called \texttt{sample\_point}, which points into the middle of the
nursery. It is initialized as:

\begin{math}
    \mathtt{sample\_point}=\mathtt{nursery\_free}+\mathtt{sample\_n\_bytes}
\end{math}

\noindent At this point there are 0 bytes allocated since the last sample, so we can
rewrite this to:

\begin{math}
    \mathtt{sample\_point}-\mathtt{nursery\_free}=\mathtt{sample\_n\_bytes}-0
\end{math}

\noindent Because \texttt{nursery\_free} gets incremented every time an allocation is
performed, the following \emph{invariant} is maintained by \texttt{allocate}:

\begin{math}
    \mathtt{sample\_point}-\mathtt{nursery\_free} = \mathtt{sample\_n\_bytes} - \mathtt{allocated}
\end{math}

\noindent From now on, we call that number `\texttt{bytes\_until\_sample}'. If
\texttt{bytes\_until\_sample} becomes negative, a sample needs to happen.
If we now set the \texttt{nursery\_limit} to the (lower) sample point instead of the 
real \texttt{nursery\_limit} (which we will call \texttt{nursery\_top} from now on), 
we can use the check \texttt{gc.nursery\_free > gc.nursery\_limit} to
check whether a sample needs to be taken. Thus, we have achieved the goal of
leaving \texttt{allocate} in its original efficient state.

\begin{figure}
  \centering
  \includegraphics[width=9cm]{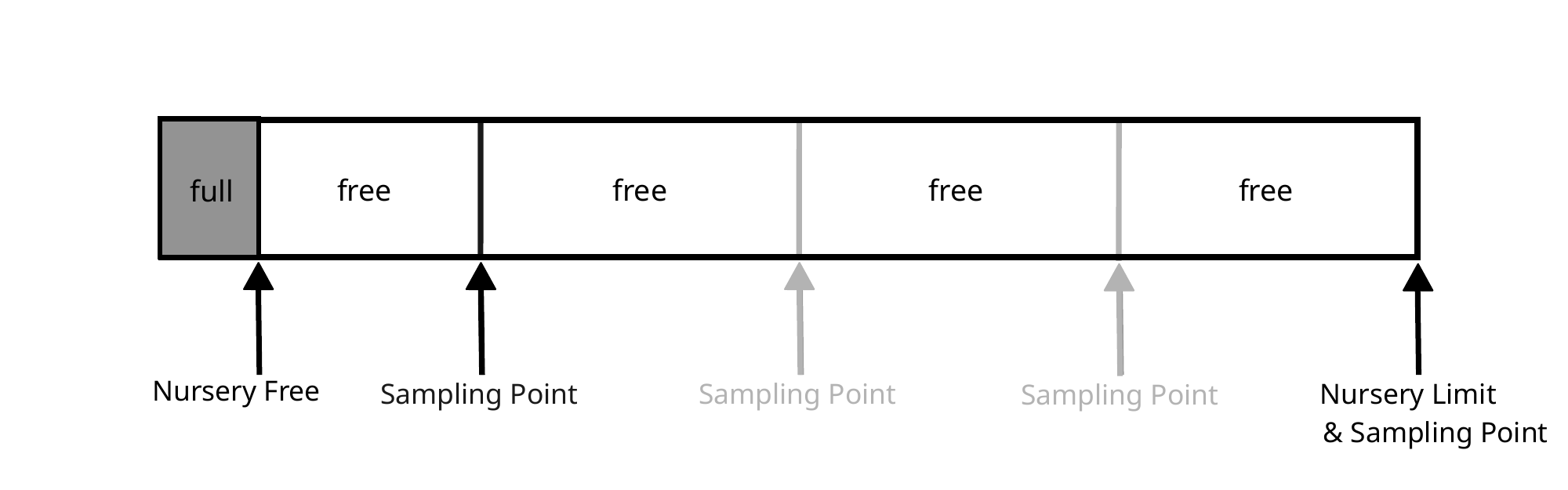}
  \caption{Nursery with Sample Points}
  \label{fig:nursery_sampling}
\end{figure}

Figure~\ref{fig:nursery_sampling} is an illustration of what the nursery with a
\texttt{sample\_point} may look like. It is a conceptual simplification as only
one \texttt{sample\_point} exists at any given time. After we created the
\texttt{sample\_point}, it will be used as \texttt{nursery\_limit} as opposed to
the actual end of the nursery.

When \texttt{nursery\_free} exceeds \texttt{nursery\_limit} by the bump-pointer
mechanism in \texttt{allocate}, this could now be for two different reasons.
Therefore,
\texttt{collect\_and\_reserve} must check whether we're really out of nursery
space and must collect the nursery or if it was a \texttt{sample\_point}, and we
need to take an allocation sample. If the latter, \texttt{collect\_and\_reserve}
will call VMProf directly to trigger a stack sample. Then it computes the next
\texttt{sample\_point} by adding \texttt{sample\_n\_bytes} and sets it as
\texttt{nursery\_limit} if it fits into the nursery. Otherwise, it sets the
\texttt{nursery\_limit} to \texttt{nursery\_top}. Figure~\ref{fig:collect_and_reserve}
shows pseudocode for these mechanisms.\footnote{
The pseudocode in Figure~\ref{fig:collect_and_reserve} is somewhat simplified.
It does not deal with the situation where an object is bigger than the sampling
period correctly. Such an object needs to be sampled more than once.}

\begin{figure}
  \begin{lstlisting}
def collect_and_reserve(size):
    # Check if we exceeded a sample point
    if gc.nursery_limit == gc.sample_point:
        # Sample and move sample_point forward
        vmprof.sample_now()
        gc.sample_point += sample_n_bytes

        # Set sample point as new nursery_limit if it fits into the nursery
        gc.nursery_limit = min(gc.sample_point, gc.nursery_top)

        # Is there enough memory left inside the nursery
        if gc.nursery_free <= gc.nursery_limit:
            # nursery_free was already incremented in allocate, thus we need to substract the object's size
            result = gc.nursery_free - size
            return result
    # Normal collect_and_reserve from here on
    ....
  \end{lstlisting}
  \caption{Pseudocode for modified \texttt{collect\_and\_reserve} function}
  \label{fig:collect_and_reserve}
\end{figure}

\subsection{Sampling period bigger than the nursery size}

\begin{figure}
  \centering
  \includegraphics[width=9cm]{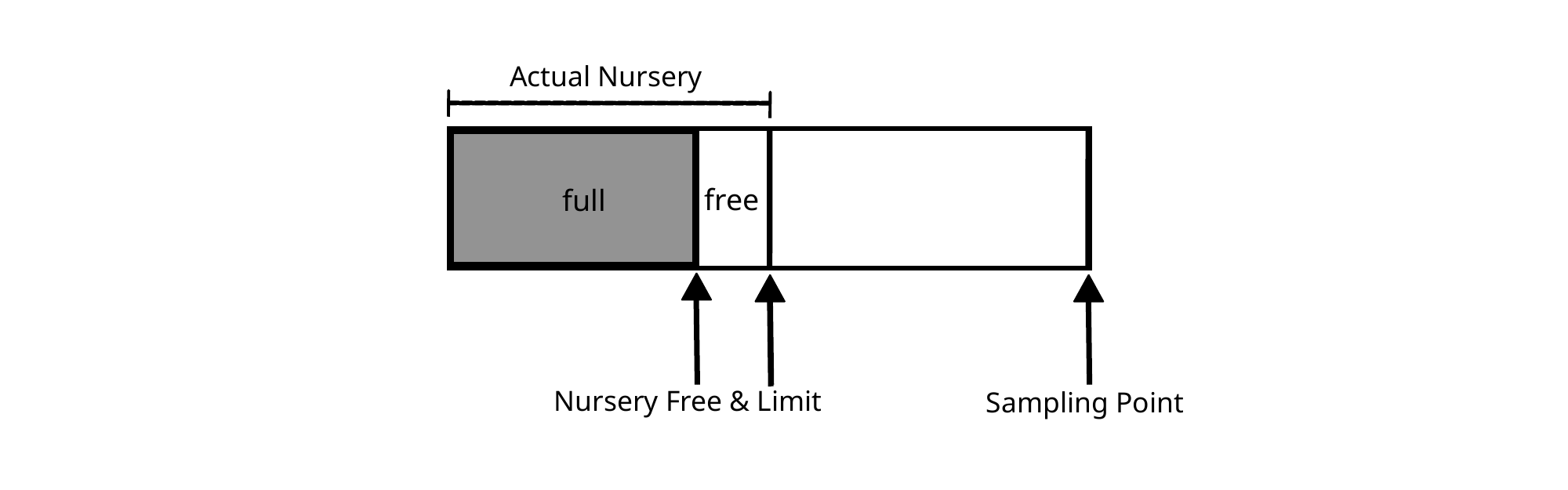}
  \caption{Sample Point outside of Nursery}
  \label{fig:larger_than_nursery}
\end{figure}

If we want to profile long-running applications, we might want to set the
\texttt{sample\_n\_bytes} parameter to values that are bigger than the size
of the nursery. In this subsection we will work through what needs to happen
when setting \texttt{sample\_point} to a point outside the nursery. If that
is the case it can't be used as \texttt{nursery\_limit}, because otherwise
\texttt{allocate} wouldn't notice that the nursery is full.
But we still need to track somehow,
after how many minor collections the samples should happen.
Figure \ref{fig:larger_than_nursery} illustrates that scenario,
\texttt{sample\_point} is outside the nursery and is therefore not used as \texttt{nursery\_limit}.

\begin{figure}
  \centering
  \includegraphics[width=9cm]{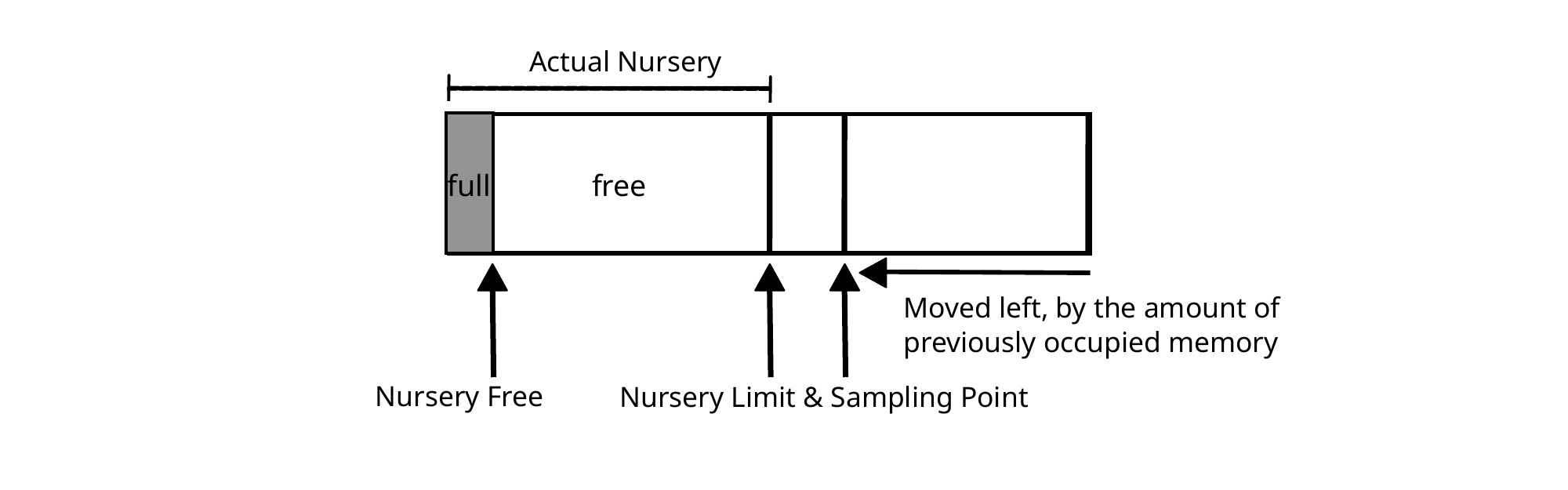}
  \caption{Sample Point moved left after Minor GC}
  \label{fig:larger_than_nursery_post_minor}
\end{figure}

To track when the next sample should happen,
we need to make sure our \texttt{bytes\_until\_sample} invariant from Section~\ref{sec:sampling_efficiently} is
maintained. Since a minor collection moves the \texttt{nursery\_free} after
evacuating the nursery, we need to move the \texttt{sample\_point} by the
difference of the old value of \texttt{nursery\_free} and the new value before
finishing the minor collection. Figure \ref{fig:larger_than_nursery_post_minor} 
shows how this effectively moves the \texttt{sample\_point} to the left. 
If that point is now inside the nursery, we will use it as the \texttt{nursery\_limit}. 
If not, the next sample happens after at least another minor collection.

\subsection{Sampling Out-of-Nursery Allocations}

In Section~\ref{section-exterenal-allocations}, we introduced large-object
allocations. Objects that are too big for the nursery will be allocated
in a separate large-object space, meaning they don't pass the nursery bump-pointer logic and thus
don't get sampled there. Big objects are a crucial part of most programs allocation behavior,
therefore it's necessary to sample them too.

We can reason about them by taking a look at the \texttt{bytes\_un\-til\_sample} invariant again. Since
\texttt{nursery\_free} doesn't move for an out-of-nursery allocation, we can make
sure that the invariant is maintained by changing the \texttt{sample\_point},
and moving that to the left by the allocated size. Then we can check whether
\texttt{sample\_point} is less than \texttt{nursery\_free} to find out whether
we need to sample the out-of-nursery allocation. If yes, we sample and then add
\texttt{sample\_n\_bytes} to \texttt{sample\_point}. In this way, the invariant
is maintained and we don't need a separate mechanism for deciding whether to
sample out-of-nursery allocations.

Figure~\ref{fig:allocate_out_of_nursery} shows pseudocode for the sampling
logic in \texttt{allocate\_out\_of\_nursery}. It is slightly simplified compared
to the real code, because the real code needs to take the possibility into
account that the allocated object is so huge, that it needs to be sampled more
than once.

Changing \texttt{allocate\_out\_of\_nursery} in this way means that
out-of-nursery allocations get slightly more expensive. However, they are rarer
and anyway significantly slower than nursery allocations, because they call a
much more complicated allocation function and do extra bookkeeping.\footnote{
Just as above, the pseudocode is somewhat simplified and does not deal with
individual allocations that are larger than the sampling period correctly.
}

\begin{figure}
  \begin{lstlisting}
def allocate_out_of_nursery(size):
    gc.sample_point -= size
    if gc.sample_point < gc.nursery_free:
        vmprof.sample_now()
        gc.sample_point += sample_n_bytes
        # Set sample point as new nursery_limit if it fits into the nursery
        gc.nursery_limit = min(gc.sample_point, gc.nursery_top)
    # now perform the allocation as usual
    ...
    return ...
  \end{lstlisting}
  \caption{Pseudocode for \texttt{allocate\_out\_of\_nursery} function}
  \label{fig:allocate_out_of_nursery}
\end{figure}

\subsection{Type and Survival Profiling}
\label{type-and-survival-profiling}

Every sampled allocation stores the call stack into the profile, to find out
where in the Python program the allocation was performed. We would also like to
store some extra information to know what kind of object was being allocated.
In addition, we want to know whether the object survived at least a single minor
collection, or whether it died as a young object.

Both of these pieces of information are not available at the point where the
stack sample is taken. The type of the object is stored only after the
allocation as part of the initialization of the object header. And the survival
information is only available at the next minor collection.

\begin{figure}
  \centering
  \includegraphics[width=9cm]{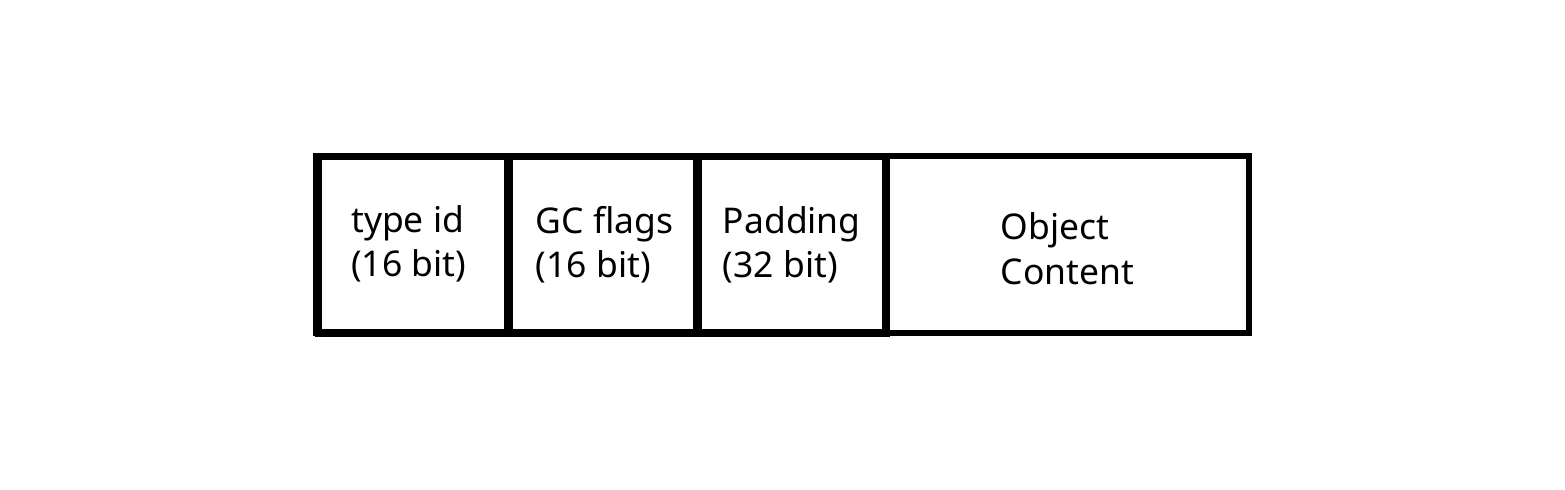}
  \caption{Object with Header and Padding}
  \label{fig:object}
\end{figure}

Every object (Figure~\ref{fig:object}) allocated by the GC has a header, composed of a 16-bit type ID and
16 bits for GC flags. (The padding is only on 64-bit platforms and omitted on 32-bit PyPy.)

In order to store the type ID and whether the object survived one minor
collection, we maintain a list of addresses of the sampled objects that were
allocated since the last minor collection (a small subset of all the allocated
objects).
At the end of the next minor collection we can walk this list of
addresses and find out whether the corresponding object survived and what its
type is. Those pieces of information are then also stored into the profile.

To be able to map type IDs of RPython types from numbers to something
human-readable, we also dump a mapping of RPython type IDs to their respective
names into the profile so that a UI tool like the vmprof-firefox-converter may
use that to display the actual type names.

Unfortunately, the type IDs correspond to the types of objects at the RPython
level, not necessarily the types of objects at the Python level. This makes
understanding the profile harder for programmers that are not knowledgeable about
the internals of PyPy. We plan to also sample the Python-level types in the
future, or to at least give more understandable names to the RPython-level type
names.

\subsection{Storing GC Statistics}

In addition to information about the sampled objects, we also store some general
information about the state of the GC heap into the profile at every minor
collection.
That information is the following:
The \texttt{total\_size\_of\_arenas} tells us how much space the GC actually has to
allocate tenured objects, while \texttt{total\_memory\_used} tells us how much of that
is already occupied. But there is more to a VM than just the memory the GC
manages, thus the \texttt{vmRSS} tells us how much memory PyPy consumes from the point
of view of the operating system.
Finally, the \texttt{GC state} tells the current major collection phase, which is one
of: \texttt{scanning, marking, sweeping, finalizing}.

\subsection{Correctness}

While developing allocation sampling, we used various techniques for
ensuring its correctness and robustness.
We started with writing simple unit tests for the allocation sampling logic.
However, PyPy's GC and the allocation sampling logic are entangled and have
complex interactions, thus we didn't trust handwritten unit tests to
sufficiently cover these interactions.
Therefore, we added support for allocation sampling into the already existing
randomized testing facility
(fuzzer)\footnote{\anonurl{\url{https://pypy.org/posts/2024/03/fixing-bug-incremental-gc.html}}},
for PyPy's GC. This randomized test uses
Hypothesis\footnote{\url{https://github.com/HypothesisWorks/hypothesis}}~\cite{MacIver_Hypothesis_A_new_2019,maciver_et_al:LIPIcs.ECOOP.2020.13},
a property-based testing framework for Python.

Fuzzing PyPy's GC with Hypothesis has two phases. The first phase is generating
random action sequences. Those actions consist of object-, string-, or array
allocations, freeing allocated objects, accessing an object and manually forcing
a minor collection. We added new actions, which are for activating and
deactivating allocation sampling with a random sampling period. In the second
phase, these actions are executed against the GC implementation and their
intermediate results asserted. If there is a bug in the GC, e.g., freeing an
object too early, the fuzzer could produce a random action sequence that leads
to an error when accessing an already freed object.

When generating these actions, we also keep track of how much memory will be
allocated when they are executed. With this information, we can decide if each
generated allocation action should trigger a sample. When these actions are
executed in the second phase, we can check for each allocation if it should
trigger a sample and if it actually did. For a failing check we then get the
sequence of actions that led to the failed check, so we can trace the bug down.

Fuzzing was very helpful for getting rid of many bugs inside the allocation
sampling logic, because it demonstrated interactions between sampling and
garbage collection that we hadn't foreseen properly.

To give a simple example, at some point in the development process, it was
possible to disable allocation sampling without enabling it first. However,
doing so led to segfaults. On disabling allocation sampling, the
\texttt{nursery\_limit} was set to \texttt{nursery\_top}. But since allocation
sampling hadn't been enabled before, \texttt{nursery\_top} was a non-initialized
pointer. This situation was generated by the fuzzer and we fixed it.

\section{vmprof-firefox-converter}
\label{sec:firefox}

The vmprof-firefox-converter is a tool for converting VMProf profiles into the firefox-processed-profile format, which can then be imported and displayed in the Firefox-Profiler.
Since the Firefox Profiler is a tool for analyzing the performance of web-based applications
some of its UI elements are JavaScript- or web-specific. One example is the
frame-filter radio buttons named 'All Frames', 'JavaScript' and 'Native' shown
in Figure \ref{fig:firefox_view}. Here we re-use the Firefox-Profiler as a UI
for VMProf; unfortunately we cannot (or at least are not aware on how to)
change the name or description of some UI elements.
The converter was adapted throughout the development of allocation-sampling for PyPy, to work with all the new information that can be extracted from PyPy's GC.
Notably, we arrange to display the allocated object type on top of the sampled
call stacks. We also show additional memory timelines for the heap statistics
discussed in Section~\ref{type-and-survival-profiling}, and we add markers for GC major collection phases.

Figure~\ref{fig:firefox_view} shows an example profile.
The profile consists of two separate sample timelines, one for time-sampling and the other for allocation-sampling.

The colorization\footnote{The colors yellow, blue, purple, and orange specify the different types of code execution.
Those code execution categories are yellow for interpreted, blue for native code (i.e., execution of a C-extension),
and finally, orange and purple for jitted code (orange hinting the transition
from interpreter to JIT).}
of those timelines represents the density of categories of the top-level frames.
The top-level frames in the `GC-Sampled'-timeline represent the sampled objects.
Green colored objects were collected while red indicates tenured objects.

The `Memory' timelines can be identified by hovering over them.
Unfortunately, we currently are not aware of how to change their names to
something other than `Memory'.

\section{Evaluation}
\label{sec:evaluation}

\subsection{Overhead}
\label{sec:evaluation_overhead}

Our most important goal of introducing the allocation sampling mechanism was to
make the overhead of sampling allocations configurable. In this section we
evaluate the overhead of sampling allocations in dependency on the sampling
period. This evaluation is preliminary, with a limited benchmark set.

The following plot shows the sampling period versus the overhead, 
which is computed as

\begin{math}
runtime\_with\_sampling / runtime\_without\_sampling
\end{math}

\begin{figure}
  \centering
  \includegraphics[width=8cm]{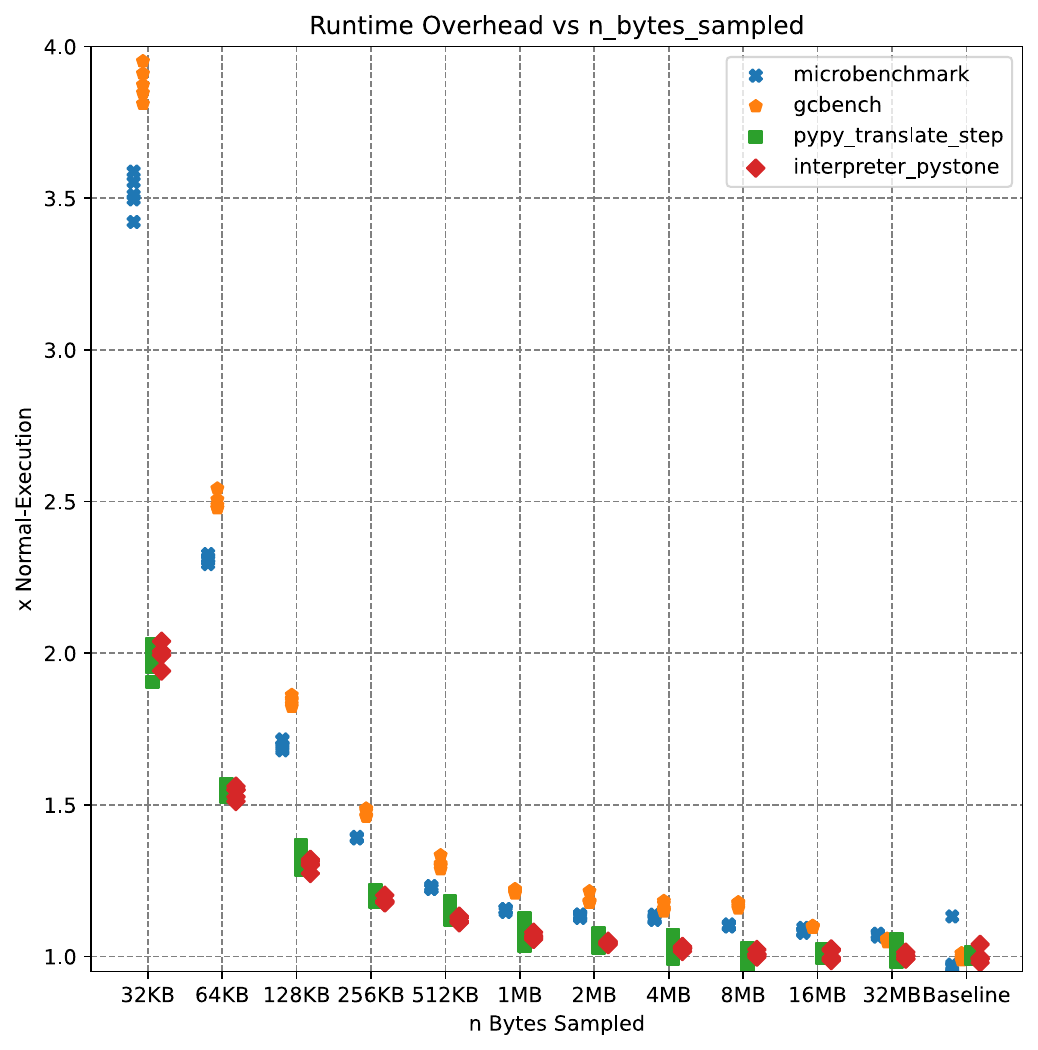}
  \caption{Allocation Sampling Period vs. Overhead}
  \label{fig:as_overhead}
\end{figure}

\begin{figure}
  \centering
  \includegraphics[width=8cm]{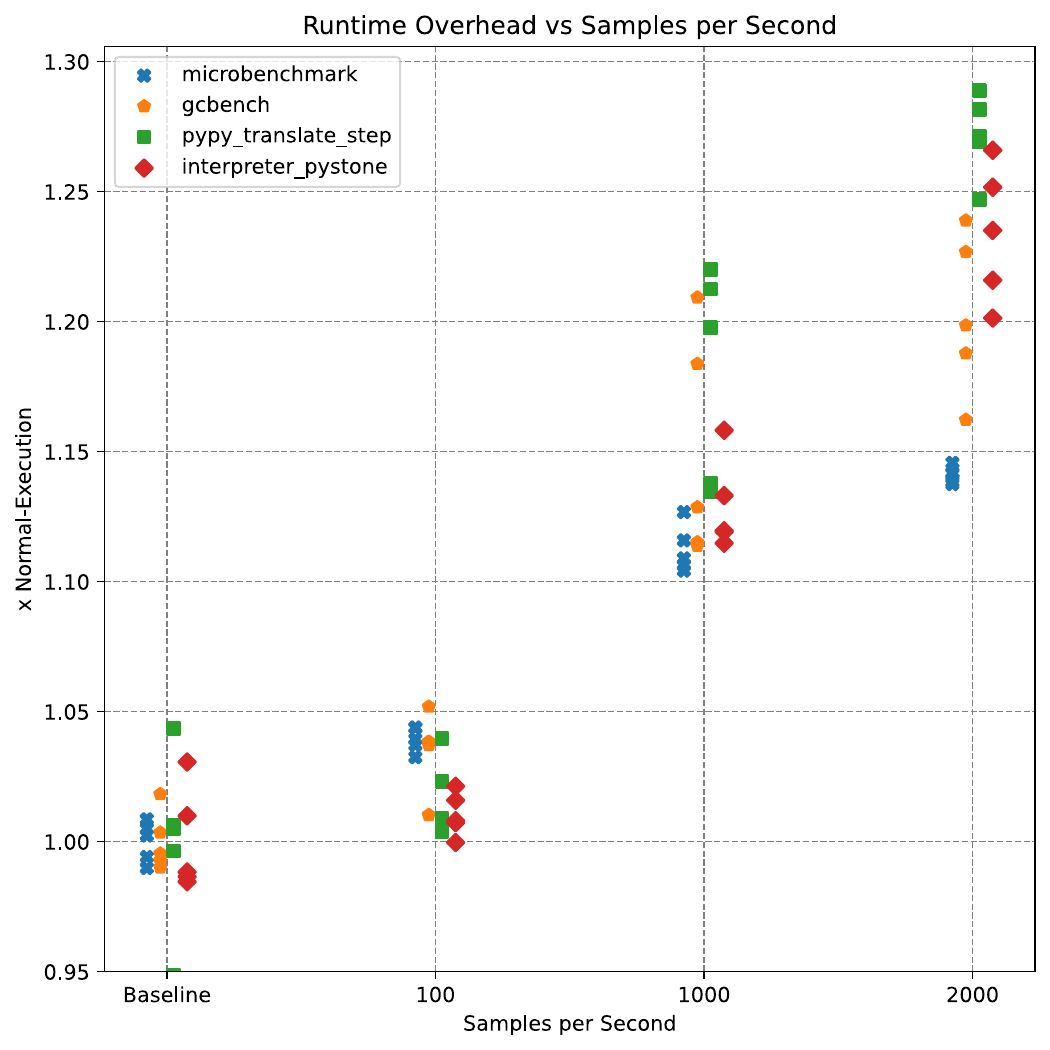}
  \caption{Time Sampling Period vs. Overhead}
  \label{fig:ts_overhead}
\end{figure}

\noindent Figure~\ref{fig:as_overhead} shows the benchmarks with allocation sampling.
We also measured the overhead of the previously existing time-based sampling in
Figure~\ref{fig:ts_overhead} as a point of comparison. All benchmarks were
executed five times on our self-built, modified PyPy with JIT and native
profiling enabled. Every dot in the plot is one run of a benchmark.

Figure~\ref{fig:as_overhead} shows that the overhead varies with the allocation sampling period.

Time-sampling with the same setup introduces overhead from a maximum of 30\% for 2000 samples per second down to a maximum of ~5\% for 100 samples per second.
Both with allocation and time sampling, it is possible to reach any amount of overhead and any level of profiling precision desired.
The best approach probably is to try out a sampling period, and choose what gives the right trade-off between precision and overhead. 

Run time, general memory usage, and memory usage per second are shown in Figure~\ref{fig:stats}.
Our benchmarks display different memory characteristics;
pypy\_translate\_step is long-running, but has the second-lowest allocation rate of only 0.580 GBs.
On the other hand gcbench, interpreter\_pystone and the microbenchmark are quite short-running, 
but two of them (microbenchmark and gcbench) have the highest allocation rates of 6.063 GB/s and 5.222 GB/s.

\begin{figure*}
  \begin{tabular}{lrrrr}
  \toprule
  Name & Run time (s) & GB Allocated & GB/s Allocated \\
  \midrule
  microbenchmark & 4.44 & 26.937 & 6.063  \\
  gcbench & 0.69 & 3.606 & 5.222 \\
  pypy\_translate\_step & 250.06 & 144.917 & 0.580  \\
  interpreter\_pystone & 8.10 & 4.494 & 0.555  \\
  \bottomrule
  \end{tabular}
  \caption{Benchmark Memory statistics without Sampling}
  \label{fig:stats}
\end{figure*}

Additionally, the average sampling rate in samples per second, overhead, and the
overhead normalized to 1000 samples per second, for 32KB allocation sampling are
shown in Figure~\ref{fig:stats2}.
The average sampling rate is computed as 

\begin{math}
    num\_samples / runtime\_in\_seconds
\end{math}

\noindent and the 1000 samples per second normalized overhead is computed as

\begin{math}
     1 + (overhead - 1) / num\_samples * 1000.
\end{math}

\noindent (The minus one results from 1 being the baseline in overhead
calculation, i.e. overhead minus 1 is the additional overhead)

\begin{figure*}
  \begin{tabular}{lrrr}
  \toprule
  Name & Samples/s & Overhead & Norm. Overhead \\
  \midrule
  microbenchmark & 52658 & 3.51 & 1.05 \\
  gcbench & 41081 & 3.88 & 1.07 \\
  pypy\_translate\_step & 8987 & 1.97 & 1.11 \\
  interpreter\_pystone & 8483 & 1.99 & 1.12 \\
  \bottomrule
  \end{tabular}
    \caption{32KB Allocation Sampling statistics}
  \label{fig:stats2}
\end{figure*}

It's visible that the seemingly high overhead at 32KB allocation sampling in Figure~\ref{fig:as_overhead}
is a result of the big number of samples per second.
Comparing the normalized 1000 samples per second allocation sampling overhead to the `real' time-sampling values in Figure~\ref{fig:ts_overhead},
one can see that allocation sampling performs slightly better. 

Another important thing to keep in mind, is the size of the resulting profile.
Sampling allocations in long-running code with a high sampling period (high period = low value for \texttt{sample\_n\_bytes}) can quickly lead to a big number of samples.
The resulting profile could consume a great amount of disk space.

For example, the \texttt{pypy\_translate\_step} benchmark with 32KB allocation sampling runs
for about 8 minutes, resulting in a profile of about 4GB in size in VMProf's binary
format (converting it to the Firefox Profiler JSON would make smaller, because
it is more efficiently organized but harder to write).
For profiling long-running programs, it is crucial to choose a sampling period
that gives the right trade-off, not just between overhead and profiling precision, but also disk space consumption.

Huge profiles could lead to the assumption that disk speed has a relevant impact on profiling overhead.
We assume that the most overhead arises from walking the Python call stack and especially from walking the C stack with \texttt{libunwind}\footnote{\url{https://github.com/libunwind/libunwind}} (native profiling).
Therefore, we cannot preclude a significant correlation between call stack depth
and stack walking overhead, but those are questions for further
research.

All benchmarks (see Section~\ref{eval-benchmarks}) executed on:

\begin{itemize}
  \item{Kubuntu 24.04 (Linux Kernel 6.8.0-60)}
  \item{AMD Ryzen 7 5700U}
  \item{24gb DDR4 3200MHz (dual channel)}
  \item{SSD benchmarking at read: 1965 MB/s, write: 227 MB/s (Sequential 1MB 1 Thread 8 Queues)}
  \item{Modified PyPy\footnote{\anonurl{\url{https://github.com/Cskorpion/pypy/tree/gc_allocation_sampling_obj_info_u_2.7}}}}  
  \item{Modified VMProf\footnote{\anonurl{\url{https://github.com/Cskorpion/vmprof-python/tree/pypy_gc_allocation_sampling_obj_info}}}}
  \item{Modified vmprof-firefox-converter\footnote{\anonurl{\url{https://github.com/Cskorpion/vmprof-firefox-converter/tree/allocation_sampling_obj_info}}}}
\end{itemize}

\subsection{Case Study: Removing Unnecessary Allocations in the PyPy JIT}

We didn't perform a serious attempt at doing a user study for the VMProf
allocation sampling tooling. However, we found some unnecessary allocations in
PyPy's JIT compiler with the tool, which we want to present here as a case
study.

\begin{figure}
  \centering
  \includegraphics[width=8cm]{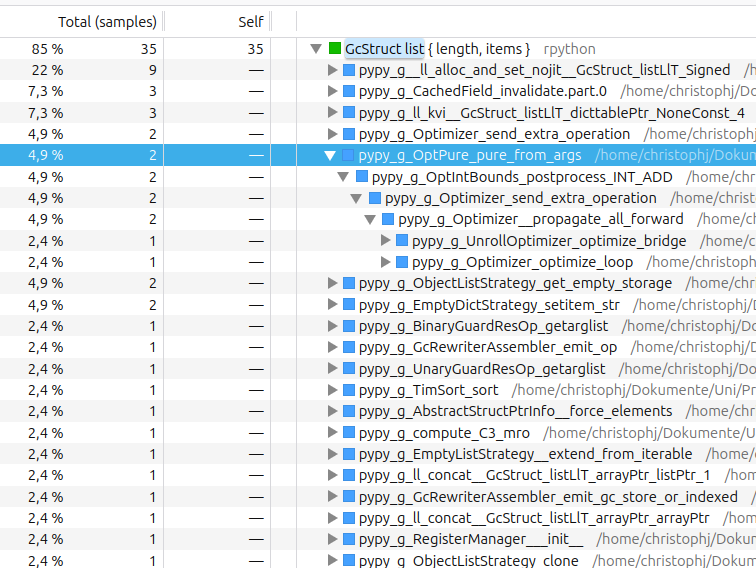}
  \caption{Firefox Profiler: Call Tree View}
  \label{fig:pure_from_args_calltree}
\end{figure}

We profiled some SymPy\footnote{\url{https://www.sympy.org}} functions with allocation sampling.
Figure~\ref{fig:pure_from_args_calltree} shows the resulting profile opened with
the Firefox Profiler's call tree view filtered for functions allocating
(RPython) lists.

There we encountered two functions from the optimizer of PyPy's JIT, \texttt{pure\_from\_args}
and \texttt{postprocess\_INT\_ADD}. Both those functions take part in optimizing
pure integer
operations.

If PyPy's JIT encounters an \texttt{INT\_ADD} operation that cannot be optimized
away\footnote{\anonurl{\url{https://pypy.org/posts/2024/10/jit-peephole-dsl.html}}}
\texttt{postprocess\_INT\_ADD} will be called to also cache some arithmetic rewrites of
that addition. E.g. if the JIT emits an operation $x = a + b$ it will remember
that $x - a$ can be optimized to $b$ from then on.
This is done by calling an API \texttt{pure\_from\_args(INT\_SUB, [arg0, arg1])}. Similar
logic exists for other integer operations like multiplication or xor.

The reason why these \texttt{postprocess\_...} functions of the JIT appear in the
memory profile is that they all allocate a list. This list is extremely
short-lived, one call level deeper its elements are read out again and the list
is then discarded. Then, one level deeper,
yet another list is built, which is discarded one function call further.
Additionally, in all of these \texttt{postprocess\_...} methods, there are never more than
two arguments passed to \texttt{pure\_from\_args} inside that list.

After seeing \texttt{pure\_from\_args} in the memory allocation profile, we decided to
rewrite it to make the list allocations unnecessary. To achieve this, we split
it up into two functions, \texttt{pure\_from\_args1} and \texttt{pure\_from\_args2}, that take
the elements of the list as extra arguments directly, foregoing the allocation.
Then all call sites were
adapted\footnote{\anonurl{\url{https://github.com/pypy/pypy/commit/ef590f639e529ebe319c7d5ff8f5e03e31bcc304}}}
to either call \texttt{pure\_from\_args1} or \texttt{pure\_from\_args2} directly, and thus saving
two list allocations per \texttt{pure\_from\_args} call.

On its own, this optimization does not yield a measurable improvement to the
warmup time of PyPy's JIT. We plan to improve the allocations in PyPy's
interpreter and JIT compiler more systematically using the new profiling tools
in the near future and hope that this work will lead to measurable results.

\section{Future Work}
\label{sec:future}

In this section we want to discuss our ideas to extend and improve our work.

While the check for whether a memory sample should be taken is free, since it
is folded into the GC allocation logic, actually doing stack samples is
relatively expensive. We inherited this logic from the existing VMProf
time-based sampling and did not try to optimize it for this paper.
Reducing the overhead per sample could be done by not walking the entire stack every
time, but mark walked stack frames so that for the next sample, we only need to
walk up to an already marked stack frame. This could reduce the overhead because
stacks typically do not change completely from sample to sample, and there is indeed a
significant correlation between stack depth and overhead.

Other directions of improvement would involve moving other sources of
information in the PyPy virtual machine into the profiles.
PyPy has a logging interface, which is used to log GC and JIT events with a
timestamp. Unfortunately those timestamps are the clock counts read from the
CPU's \texttt{TSC} (Time Stamp Counter, number of cycles since last reset) register
(at least on x86/x86\_64), which are not perfectly suitable for measuring time.
Our modified version of VMProf on the other hand uses timestamps retrieved with
Unix' \texttt{CLOCK\_MONOTONIC}. This means we cannot exactly associate log events
with VMProf samples. An easy fix would be to use the same timestamps for
the logging facility as we do for VMProf, but this might introduce too much overhead. A better
way of associating them, could be to record the \texttt{TSC} with each sample so we'd
get an approximate alignment of logged events and samples.

Another potential idea for future improvements would be using \texttt{ptrace}
to trace system calls. That could give an insight on where/when/how much the executed
code spent time opening files, reading from files, waiting for subprocesses to
finish, etc.

We are also hoping to transfer some of the techniques used here for profiling
PyPy to profile other RPython languages, such as the CPU simulators generated
by Pydrofoil~\cite{bolztereick2025pydrofoil}.

\section{Related Work}
\label{sec:related}

Finally, we want to discuss related approaches for memory profiling, allocation-
and object lifetime sampling.

Jump et al.~\cite{jump_blackburn_mckinley_2004} employ similar bump-pointer sampling logic but use that for sampling allocation-site lifetime behavior for dynamic pretenuring in a Java VM.
We sample the call stack for general performance analysis, but also record (a more primitive) object lifetime measure.
Their goal is to reduce GC load through analyzing allocation-site lifetimes and directly allocating objects from long-living allocation sites to the old-space.
Their results show a reduction of GC time of up to two times in some benchmarks, but also a performance degradation in other benchmarks.
We implement a mechanic for analyzing performance, that does not directly increase performance, furthermore introduces (fully adjustable) overhead to PyPy when used.

Harris~\cite{harris_2000} uses a slightly different approach. He samples for allocation site and class lifetime statistics for dynamic pretenuring in Java.
In contrast to our approach, Harris takes a sample every time a thread's LAB (small memory area for one single thread) overflows.
Then on every first generation collection, it is recorded if the objects got tenured or died.
Another key difference is that he also keeps track of when an object died inside the second generation heap,
while we don't care for objects after they left the nursery.

Poireau~\cite{poireau} follows a more low-level approach.
They intercept calls to \texttt{malloc/calloc/free/realloc} and use perf to take a sample with a certain probability for each allocated byte.
Their target is to use the collected samples for debugging and leak detection, more than memory housekeeping optimizations and general performance analysis.
A key difference to our work is that intercepting calls to \texttt{malloc} etc. works with any language using malloc etc.,
while our approach is limited to PyPy and (theoretically) any other RPython language with VMProf support that uses the same GC implementation as PyPy.

Memray~\cite{memray} on the other hand, is an instrumenting memory profiler for
CPython.
That means instead of sampling, Memray records every allocation; thus, it introduces non-adjustable overhead.
Memray has the same use cases as our approach to allocation sampling for PyPy. 
That is to help developers understand where their code allocates memory, find leaks, and discover optimization potential. 

Guile, the GNU Scheme implementation, has a statistical memory profiler. It
works by taking a stack sample every time the garbage collector
runs.\footnote{\url{https://www.gnu.org/software/guile/manual/html\_node/Statprof.html\#index-gcprof}}
Since a GC run is likely to be triggered by functions that allocate a lot, this
effectively samples functions with high allocation rates. This approach can be
seen as a special case of what we did in this paper, it corresponds to setting
\texttt{sample\_n\_bytes} to the size of the nursery so that minor collection
and sampling always happen together.

In contrast to Memray and Poireau, Sciagraph~\cite{sciagraph} is a commercial Python profiler and not open source.
Sciagraph does both performance and memory profiling and introduces low overhead (below 5\%)
by taking about 20 stack samples per second and sampling calls to \texttt{malloc}.
As allocation sampling for PyPy with VMProf can also be combined with time sampling, its use case is the same as Sciagraph's.
However, Sciagraph's main use case is to find memory leaks and to find out which
allocation sites contribute to an application's peak memory
usage.\footnote{Itamar Turner-Trauring, personal communication: \url{https://hachyderm.io/@itamarst/114546578271860025}}

Finally, Burchell et al.~\cite{burchell_2024} use late compiler phase instrumentation to profile the Java GraalVM.
They insert instrumentation code only at the latest possible point in time during JIT-compilation;
thus only code that is jitted will be profiled.
In contrast, our approach is sampling everything that allocates memory, which the JIT itself does a lot.
Burchell et al. emphasize the problem of inlining in profiling,
that is, functions not appearing in the profile because they've been inlined into the caller.
Their approach is to estimate the amount of run time that is spent in an inlined function inside its caller.
VMProf on the other hand, handles this by maintaining detailed information about
inlining for every point in the generated machine code, such that Python stacks
can be precisely reconstructed by the profiler.

\section{Conclusion}
\label{sec:conclusion}

In this paper, we introduced allocation sampling in PyPy's GC with VMProf.
Using this tool to simultaneously do allocation and time sampling can give insight into where the
program spends time, and what functions allocate much memory, leading to garbage collections.
This tool is aimed at both PyPy developers and non-PyPy developers, 
with the target of being easy to use while introducing little overhead.
Right now, the tool is still in development, 
we hope to merge and release it with a PyPy release some time soon.

\paragraph{Acknowledgements} We'd like to thank Matti Picus for his work on the
PyPy and VMProf projects. Also, Max Bernstein and Andy Wingo have provided
helpful comments for earlier drafts of this paper.

\section{Benchmarks}

\label{eval-benchmarks}

\begin{itemize}
  \item microbenchmark 
    \begin{itemize}
      \item \footnote{\anonurl{\url{https://github.com/Cskorpion/microbenchmark}}}
      \item \textit{pypy microbench.py 65536}
    \end{itemize}

    \item gcbench
    \begin{itemize}
      \item github\footnote{\url{https://github.com/pypy/pypy/blob/main/rpython/translator/goal/gcbench.py}}
      \item print statements removed
      \item \textit{pypy gcbench.py 1}
    \end{itemize}

    \item pypy translate step
    \begin{itemize}
      \item first step of the pypy translation (annotation step)
      \item \textit{pypy path/to/rpython --opt=0 --cc=gcc --dont-write-c-files --gc=incminimark --annotate path/to/pypy/goal/targetpypystandalone.py }
    \end{itemize}
  
    \item interpreter pystone 
    \begin{itemize}
      \item pystone benchmark on top of an interpreted pypy on top of a translated pypy
      \item \textit{pypy path/to/pypy/bin/pyinteractive.py -c "import test.pystone; test.pystone.main(1)"}
    \end{itemize}
\end{itemize}

\bibliographystyle{ACM-Reference-Format}
\bibliography{allocation-sampling}


\begin{thebibliography}{15}


\ifx \showCODEN    \undefined \def \showCODEN     #1{\unskip}     \fi
\ifx \showISBNx    \undefined \def \showISBNx     #1{\unskip}     \fi
\ifx \showISBNxiii \undefined \def \showISBNxiii  #1{\unskip}     \fi
\ifx \showISSN     \undefined \def \showISSN      #1{\unskip}     \fi
\ifx \showLCCN     \undefined \def \showLCCN      #1{\unskip}     \fi
\ifx \shownote     \undefined \def \shownote      #1{#1}          \fi
\ifx \showarticletitle \undefined \def \showarticletitle #1{#1}   \fi
\ifx \showURL      \undefined \def \showURL       {\relax}        \fi
\providecommand\bibfield[2]{#2}
\providecommand\bibinfo[2]{#2}
\providecommand\natexlab[1]{#1}
\providecommand\showeprint[2][]{arXiv:#2}

\bibitem[Ancona et~al\mbox{.}(2007)]%
        {AnconaD:rpystetrdastol}
\bibfield{author}{\bibinfo{person}{Davide Ancona}, \bibinfo{person}{Massimo
  Ancona}, \bibinfo{person}{Antonio Cuni}, {and} \bibinfo{person}{Nicholas~D.
  Matsakis}.} \bibinfo{year}{2007}\natexlab{}.
\newblock \showarticletitle{{RPython: a step towards reconciling dynamically
  and statically typed OO languages}}. In \bibinfo{booktitle}{\emph{Proceedings
  of the 2007 Symposium on Dynamic Languages}} (Montreal, Quebec, Canada)
  \emph{(\bibinfo{series}{DLS '07})}. \bibinfo{publisher}{Association for
  Computing Machinery}, \bibinfo{address}{New York, NY, USA},
  \bibinfo{pages}{53–64}.
\newblock
\showISBNx{9781595938688}
\href{https://doi.org/10.1145/1297081.1297091}{doi:\nolinkurl{10.1145/1297081.1297091}}


\bibitem[backtrace labs(2020)]%
        {poireau}
\bibfield{author}{\bibinfo{person}{backtrace labs}.}
  \bibinfo{year}{2020}\natexlab{}.
\newblock \bibinfo{booktitle}{\emph{Poireau GitHub}}.
\newblock
\urldef\tempurl%
\url{https://github.com/backtrace-labs/poireau}
\showURL{%
\tempurl}


\bibitem[bloomberg(2022)]%
        {memray}
\bibfield{author}{\bibinfo{person}{bloomberg}.}
  \bibinfo{year}{2022}\natexlab{}.
\newblock \bibinfo{booktitle}{\emph{memray GitHub}}.
\newblock
\urldef\tempurl%
\url{https://github.com/bloomberg/memray}
\showURL{%
\tempurl}


\bibitem[Bolz et~al\mbox{.}(2011)]%
        {BolzCF:allrembpeiatj}
\bibfield{author}{\bibinfo{person}{Carl~Friedrich Bolz},
  \bibinfo{person}{Antonio Cuni}, \bibinfo{person}{Maciej Fijałkowski},
  \bibinfo{person}{Michael Leuschel}, \bibinfo{person}{Samuele Pedroni}, {and}
  \bibinfo{person}{Armin Rigo}.} \bibinfo{year}{2011}\natexlab{}.
\newblock \showarticletitle{{Allocation removal by partial evaluation in a
  tracing JIT}}. In \bibinfo{booktitle}{\emph{Proceedings of the 20th ACM
  SIGPLAN Workshop on Partial Evaluation and Program Manipulation}} (Austin,
  Texas, USA) \emph{(\bibinfo{series}{PEPM '11})}.
  \bibinfo{publisher}{Association for Computing Machinery},
  \bibinfo{address}{New York, NY, USA}, \bibinfo{pages}{43–52}.
\newblock
\showISBNx{9781450304856}
\href{https://doi.org/10.1145/1929501.1929508}{doi:\nolinkurl{10.1145/1929501.1929508}}


\bibitem[Bolz et~al\mbox{.}(2009)]%
        {BolzCF:trametlptjc}
\bibfield{author}{\bibinfo{person}{Carl~Friedrich Bolz},
  \bibinfo{person}{Antonio Cuni}, \bibinfo{person}{Maciej Fijałkowski}, {and}
  \bibinfo{person}{Armin Rigo}.} \bibinfo{year}{2009}\natexlab{}.
\newblock \showarticletitle{{Tracing the meta-level: PyPy's tracing JIT
  compiler}}. In \bibinfo{booktitle}{\emph{Proceedings of the 4th Workshop on
  the Implementation, Compilation, Optimization of Object-Oriented Languages
  and Programming Systems}} (Genova, Italy) \emph{(\bibinfo{series}{ICOOOLPS
  '09})}. \bibinfo{publisher}{Association for Computing Machinery},
  \bibinfo{address}{New York, NY, USA}, \bibinfo{pages}{18–25}.
\newblock
\showISBNx{9781605585413}
\href{https://doi.org/10.1145/1565824.1565827}{doi:\nolinkurl{10.1145/1565824.1565827}}


\bibitem[Bolz-Tereick et~al\mbox{.}(2025)]%
        {bolztereick2025pydrofoil}
\bibfield{author}{\bibinfo{person}{Carl~Friedrich Bolz-Tereick},
  \bibinfo{person}{Luke Panayi}, \bibinfo{person}{Ferdia McKeogh},
  \bibinfo{person}{Tom Spink}, {and} \bibinfo{person}{Martin Berger}.}
  \bibinfo{year}{2025}\natexlab{}.
\newblock \bibinfo{title}{Pydrofoil: accelerating Sail-based instruction set
  simulators}.
\newblock
\showeprint{arXiv:2503.04389}
\newblock
\shownote{To appear in ECOOP 2025}.


\bibitem[Burchell et~al\mbox{.}(2024)]%
        {burchell_2024}
\bibfield{author}{\bibinfo{person}{Humphrey Burchell}, \bibinfo{person}{Octave
  Larose}, {and} \bibinfo{person}{Stefan Marr}.}
  \bibinfo{year}{2024}\natexlab{}.
\newblock \showarticletitle{Towards Realistic Results for Instrumentation-Based
  Profilers for JIT-Compiled Systems}. In \bibinfo{booktitle}{\emph{Proceedings
  of the 21st ACM SIGPLAN International Conference on Managed Programming
  Languages and Runtimes}} (Vienna, Austria) \emph{(\bibinfo{series}{MPLR
  2024})}. \bibinfo{publisher}{Association for Computing Machinery},
  \bibinfo{address}{New York, NY, USA}, \bibinfo{pages}{82–89}.
\newblock
\showISBNx{9798400711183}
\href{https://doi.org/10.1145/3679007.3685058}{doi:\nolinkurl{10.1145/3679007.3685058}}


\bibitem[Harris(2000)]%
        {harris_2000}
\bibfield{author}{\bibinfo{person}{Timothy~L. Harris}.}
  \bibinfo{year}{2000}\natexlab{}.
\newblock \showarticletitle{Dynamic adaptive pre-tenuring}. In
  \bibinfo{booktitle}{\emph{Proceedings of the 2nd International Symposium on
  Memory Management}} (Minneapolis, Minnesota, USA)
  \emph{(\bibinfo{series}{ISMM '00})}. \bibinfo{publisher}{Association for
  Computing Machinery}, \bibinfo{address}{New York, NY, USA},
  \bibinfo{pages}{127–136}.
\newblock
\showISBNx{1581132638}
\href{https://doi.org/10.1145/362422.362476}{doi:\nolinkurl{10.1145/362422.362476}}


\bibitem[Jones et~al\mbox{.}(2023)]%
        {jones_garbage_2023}
\bibfield{author}{\bibinfo{person}{Richard Jones}, \bibinfo{person}{Antony
  Hosking}, {and} \bibinfo{person}{Eliot Moss}.}
  \bibinfo{year}{2023}\natexlab{}.
\newblock \bibinfo{booktitle}{\emph{The {Garbage} {Collection} {Handbook}:
  {The} {Art} of {Automatic} {Memory} {Management}} (\bibinfo{edition}{2nd
  edition} ed.)}.
\newblock \bibinfo{publisher}{Chapman \& Hall/CRC}, \bibinfo{address}{Boca
  Raton}.
\newblock
\showISBNx{978-1-03-221803-8}


\bibitem[Jump et~al\mbox{.}(2004)]%
        {jump_blackburn_mckinley_2004}
\bibfield{author}{\bibinfo{person}{Maria Jump}, \bibinfo{person}{Stephen~M.
  Blackburn}, {and} \bibinfo{person}{Kathryn~S. McKinley}.}
  \bibinfo{year}{2004}\natexlab{}.
\newblock \showarticletitle{Dynamic object sampling for pretenuring}. In
  \bibinfo{booktitle}{\emph{Proceedings of the 4th International Symposium on
  Memory Management}} (Vancouver, BC, Canada) \emph{(\bibinfo{series}{ISMM
  '04})}. \bibinfo{publisher}{Association for Computing Machinery},
  \bibinfo{address}{New York, NY, USA}, \bibinfo{pages}{152–162}.
\newblock
\showISBNx{1581139454}
\href{https://doi.org/10.1145/1029873.1029892}{doi:\nolinkurl{10.1145/1029873.1029892}}


\bibitem[MacIver and Donaldson(2020)]%
        {maciver_et_al:LIPIcs.ECOOP.2020.13}
\bibfield{author}{\bibinfo{person}{David~R. MacIver} {and}
  \bibinfo{person}{Alastair~F. Donaldson}.} \bibinfo{year}{2020}\natexlab{}.
\newblock \showarticletitle{{Test-Case Reduction via Test-Case Generation:
  Insights from the Hypothesis Reducer}}. In \bibinfo{booktitle}{\emph{34th
  European Conference on Object-Oriented Programming (ECOOP 2020)}}
  \emph{(\bibinfo{series}{Leibniz International Proceedings in Informatics
  (LIPIcs)}, Vol.~\bibinfo{volume}{166})},
  \bibfield{editor}{\bibinfo{person}{Robert Hirschfeld} {and}
  \bibinfo{person}{Tobias Pape}} (Eds.). \bibinfo{publisher}{Schloss Dagstuhl
  -- Leibniz-Zentrum f{\"u}r Informatik}, \bibinfo{address}{Dagstuhl, Germany},
  \bibinfo{pages}{13:1--13:27}.
\newblock
\showISBNx{978-3-95977-154-2}
\showISSN{1868-8969}
\href{https://doi.org/10.4230/LIPIcs.ECOOP.2020.13}{doi:\nolinkurl{10.4230/LIPIcs.ECOOP.2020.13}}


\bibitem[MacIver et~al\mbox{.}(2019)]%
        {MacIver_Hypothesis_A_new_2019}
\bibfield{author}{\bibinfo{person}{David~R. MacIver}, \bibinfo{person}{Zac
  Hatfield-Dodds}, {and} \bibinfo{person}{{many other contributors}}.}
  \bibinfo{year}{2019}\natexlab{}.
\newblock \showarticletitle{{Hypothesis: A new approach to property-based
  testing}}.
\newblock  (\bibinfo{date}{Nov.} \bibinfo{year}{2019}).
\newblock
\href{https://doi.org/10.21105/joss.01891}{doi:\nolinkurl{10.21105/joss.01891}}


\bibitem[Rigo and Pedroni(2006)]%
        {rigo_pypys_2006}
\bibfield{author}{\bibinfo{person}{Armin Rigo} {and} \bibinfo{person}{Samuele
  Pedroni}.} \bibinfo{year}{2006}\natexlab{}.
\newblock \showarticletitle{{PyPy}'s approach to virtual machine construction}.
  In \bibinfo{booktitle}{\emph{{DLS}}}. \bibinfo{publisher}{ACM},
  \bibinfo{address}{Portland, Oregon, USA}.
\newblock
\showISBNx{1-59593-491-X}
\href{https://doi.org/10.1145/1176617.1176753}{doi:\nolinkurl{10.1145/1176617.1176753}}


\bibitem[Turner-Trauring(2024)]%
        {sciagraph}
\bibfield{author}{\bibinfo{person}{Itamar Turner-Trauring}.}
  \bibinfo{year}{2024}\natexlab{}.
\newblock \bibinfo{booktitle}{\emph{Sciagraph homepage}}.
\newblock
\urldef\tempurl%
\url{https://www.sciagraph.com/docs/reference/limitations/}
\showURL{%
\tempurl}


\bibitem[Wilson(1992)]%
        {wilson_uniprocessor_1992}
\bibfield{author}{\bibinfo{person}{Paul~R. Wilson}.}
  \bibinfo{year}{1992}\natexlab{}.
\newblock \showarticletitle{Uniprocessor {Garbage} {Collection} {Techniques}}.
  In \bibinfo{booktitle}{\emph{Proceedings of the {International} {Workshop} on
  {Memory} {Management}}}. \bibinfo{publisher}{Springer-Verlag},
  \bibinfo{pages}{1--42}.
\newblock
\showISBNx{3-540-55940-X}
\urldef\tempurl%
\url{http://portal.acm.org/citation.cfm?id=664824}
\showURL{%
\tempurl}


\end{thebibliography}


\end{document}